# A study of long-range order in certain two-dimensional frustrated lattices

Uma Bhaumik[1,*] and A. Taraphder[2,**]

[1]Vidyasagar College, 39 Sankar Ghosh Lane, Calcutta 700006 India
[2]Department of Physics & Centre for Theoretical Studies, Indian Institute of Technology Kharagpur 721302 India

We have studied Heisenberg Antiferromagnets on two-dimensional frustrated lattices - triangular and Kagomé lattices, using linear spin-wave theory. A collinear ground state ordering is possible if one of the three bonds in each triangular plaquette of the lattice becomes weaker or frustrated. We study spiral order in the Heisenberg model along with Dzyaloshinskii-Moriya interaction and in presence of a magnetic field. The quantum corrections to the ground state energy and sublattice magnetization are calculated analytically in the case of triangular lattice with nearest neighbor interaction. The corrections depend on the DM interaction strength and the magnetic field. We find that the DM interaction stabilizes the long-range order, reducing the effect of quantum fluctuation. Similar conclusions are reached for the Kagomé lattice. We work out the linear spin-wave theory at first with only the nearest-neighbour terms for the Kagome lattice. We find that the near-neighbour interaction is not sufficient to remove the effects of low energy fluctuations. The flat branch in the excitation spectrum becomes dispersive on inclusion of further neighbor interactions. The ground state energy and the excitation spectrum have been obtained for various cases.

## I. Introduction.

In the last few decades, geometrically frustrated antiferromagnets (AFM) have become a very important subject for both experimental and theoretical research [1]. The most extensively studied systems in this category are the AFM on triangular and kagomé lattices in two dimensions and pyrochlore lattice in three dimensions [2]. A large number of studies have been devoted to the triangular lattice. P. W. Anderson [3] first proposed that the triangular lattice Heisenberg antiferromagnet has a spin-disordered ground state, similar to the frustrated square lattices with further neighbour exchange. The resonating valence bond (RVB) state is one of the possible candidates for the ground state in this regime. An estimate of ground state energy has been obtained from various RVB-type variational wave functions [4,5] and from the variational estimates of Huse and Elser [6]. But several other methods, such as spin wave theory [7,8,9], variational calculations [10], exact diagonalization of small clusters [11] and Monte Carlo [12] numerical method have indicated the possibility of a long-range order (LRO) with the ground state energy lower than the spin disordered states. The sublattice magnetization is, however, reduced considerably (~0.239) from its classical values (~0.5) due to quantum fluctuations. It is generally believed that the frustrated triangular lattice Heisenberg antiferromanet (HAFM) is quite similar to the square lattice i.e., a ground state with long-range order exists in both cases. The triangular quantum antiferromagnet (QAFM) is thought to exhibit the well-known long-range 120° order at T=0. There is no experimental evidence in favor of such a conclusion so far though. Experimental realizations of triangular lattice HAFM are materials like $VCl_2$, $VBr_2$, $NaNiO_2$ etc. Possible ground state orderings in the quantum antiferromagnets include Néel, helical, spin-liquid, spin-nematic, dimer or chiral liquid. Dimer ordering is found for various

SU(n) models for large n and may even survive in the n→2 limit, i.e., for S=1/2, for models with competing further neighbor interactions. Indeed there is no experimental realization of LRO or spin-liquid ground states. The layered insulating magnet $Cs_2CuCl_4$ [13], which is supposed to have some long-range order at very low temperature ($T_N \sim 0.62^o$K), can be described by HAFM on a triangular lattice with additional anisotropic interaction - the Dzyaloshinskii-Moriya (DM) interaction. The DM interaction produces canting between the spins, and as a result weak ferromagnetism in the AFM phase develops and staggered magnetization reduces.

Another example of a frustrated lattice in two dimensions is the kagomé lattice, consisting of corner-sharing triangles. This is a highly frustrated system with low coordination number. Though there is no experimental evidence for spin-liquid or the dimer ordered ground states so far, it is expected to have [14,15,16, 24] a spin-liquid ground state and hence no particular long-range ordering is favored. In such cases, the ground state manifold is expected to have a large number of nearly degenerate states. Physical examples of kagomé lattice includes second layer $^3$He on graphite, jarosites, $MFe_3(OH)_6(SO_4)_2$ (M = $H_3O$,Na,K,Rb,Ag etc.), $SrCr_{8-x}Ga_{4+x}O_{19}$. Recently some magnetic structures of Fe and Cr jarosites are explored, and in order to explain the low temperature behavior of these compounds, it was proposed [17] that DM interaction may be present there. The low temperature magnetic structure is a long-range ordered state where all the spins have the same component in the direction perpendicular to the kagomé plane, giving rise to a weak ferromagnetism. A recently developed system showing large magnetoresistance, the doped $GdI_2$, [18] turns out to be another example of a triangular lattice with several magnetic phases. It has a complicated magnetic phase-diagram showing broad transitions between ferro and antiferromagnetic phases and possible regions of disorder, short-range order or phase separation [19].

It is apparent that there is a lot of controversy about the ground state of such frustrated lattices. Most of the studies done in these lattices are with Heisenberg Hamiltonian only. In view of that, we have undertaken to study the Heisenberg model along with anisotropic DM interaction on both triangular and kagomé lattices. In the present work we show that anisotropic interactions like DM interaction stabilizes the LRO. It has been observed that the effect of quantum fluctuation in the ground state energy and sublattice magnetization decreases (i.e., the ground state energy decreases and the sublattice magnetization increases), as the strength of DM interaction increases. In particular, we have studied the effect of this anisotropic DM interaction on the ground state energy, sublattice magnetization and gap in the excitation spectrum. Effect of external magnetic field has also been taken into account. This paper is organized as follows. In section II, we write the general Hamiltonian and investigate the possible ground states in the triangular and kagomé lattice. In section III we shall study the collinear Néel ordering for the Heisenberg Hamiltonian only for these frustrated lattices in the ground state using spin wave theory. In section IV, we shall study the above-mentioned lattices with spiral ordered ground states. Then, in the last section, we shall conclude with our results.

## II. **The General Hamiltonian**.

The most general spin Hamiltonian for two neighboring spin-1/2 localized magnetic ions is given by

$$H_{ij} = J_{ij} S_i \cdot S_j + D_{ij} \cdot S_i \times S_j + S_i \cdot \overleftrightarrow{A}_{ij} \cdot S_j + hS_i + hS_j \qquad (1)$$

The first term is the usual Heisenberg term. The second term is the anisotropic DM interaction and the third term involving $\overset{\leftrightarrow}{A}_{ij}$ is the anisotropic symmetric exchange interaction [20]. The last term is the Zeeman term in a magnetic field. Here, the consequences of the DM interaction on the low temperature magnetic structures are explored in the case where $\boldsymbol{D}$ is perpendicular to the lattice plane. In a magnetic insulator, in terms of the super-exchange mechanism, the isotropic exchange $J_{ij}$ is proportional to $t_{ij}^2/U$ where $t_{ij}$ is the inter-site hopping and $U$ is the onsite Coulomb repulsion between electrons. It was shown by Moriya that $|\boldsymbol{D}_{ij}|$ is proportional to $|\lambda t_{ij}/\Delta U|$ where $\lambda$ is the spin-orbit coupling and $\Delta$ is the crystal field splitting and $\boldsymbol{A}_{ij}$ is proportional to $(\lambda^2 t_{ij}/\Delta^2 U)$. The third term, being one order of magnitude smaller than DM interaction can, therefore, be neglected. In addition a magnetic field in the plane is introduced. Finally, under such approximations, we take the Hamiltonian as,

$$H = J\sum_{<i,j>} S_i \cdot S_j + D \cdot \sum_{<i,j>} S_i \times S_j + h\sum_i S_i \qquad (2)$$

We have studied the triangular and kagomé lattices using this Hamiltonian. At first the classical ground states have been investigated and then the quantum corrections are obtained from a spin-wave analysis. Let us, at the outset, find the classical ground state for the above Hamiltonian in different cases.

## Classical Ground State

In general the QAFM in more than one dimensions like triangular, kagomé, pyrochlore lattices would appear to be highly frustrated. At classical level these systems are highly degenerate and have non-zero entropy at T=0 and hence no long-range order is favored. In general, at the classical level (equivalently, $S\rightarrow\infty$) the spins are assumed to be classical vectors and the spin-spin interaction energy is minimized with respect to some parameters (like the angles between the spins) for the Heisenberg Hamiltonian. In this situation several states are obtained as the possible ground states. In two-dimensional frustrated lattices the well-known candidates having long-range order are the collinear ordering and the spiral ordering. In the following we determine both the orderings classically.

A triangular plaquette is the basic building block for the two-dimensional frustrated lattices. Let us consider a single such plaquette oriented in the X-Z plane and set $h=0$ for now. In the classical limit the spins can be treated as classical vectors. Let the three spin vectors make angles $0, \theta_1, \theta_2$ and interact with each other through bonds of strengths $(J, J, \alpha J)$ respectively. Bhaumik and Bose [21] have shown for the HAFM that a single triangular plaquette, in which one of the three bonds is weaker $(\alpha < 1)$, may have several possible orderings depending upon the strength of the weaker bond. For the strength of the frustrated bond less than 0.5 (i.e., $\alpha < 0.5$), collinear ordering is possible in the classical regime (for both the collinear and spiral order, $\theta_1 = 2\theta_2$). Beyond that, the angle between the spins changes as $\theta_2 = \cos^{-1}(1/2\alpha)$ so that in the limit $\alpha = 1$ it becomes 120º Néel ordered. As an interesting example for collinear order in the triangular lattice, the row model is studied. In this model, a triangular lattice with bonds along a particular direction was assumed to be frustrated. As a result, in the limit $\alpha = 0$ it becomes a depleted square lattice. Using spin-wave theory it was established that quantum fluctuations are not strong enough to destroy the order for $\alpha < 0.32$. Keeping one of the bonds weaker we can construct a number of collinear ordered ground states in the triangular and the kagomé lattice as shown in Fig.1. The particular collinear ordering shown in Fig. 1 in the triangular lattice is revisited here because similar states appear in the phase diagram of the Hubbard model on a triangular lattice [22] and also in the study of the correlated double-exchange model on a triangular lattice [19]. A collinear order on the kagomé lattice is also considered as another example of such a frustrated lattice. In the limit $\alpha = 0$ the kagomé lattice becomes a decorated and depleted ferromagnetic square lattice with opposite spins sitting at the midpoint of the bonds.

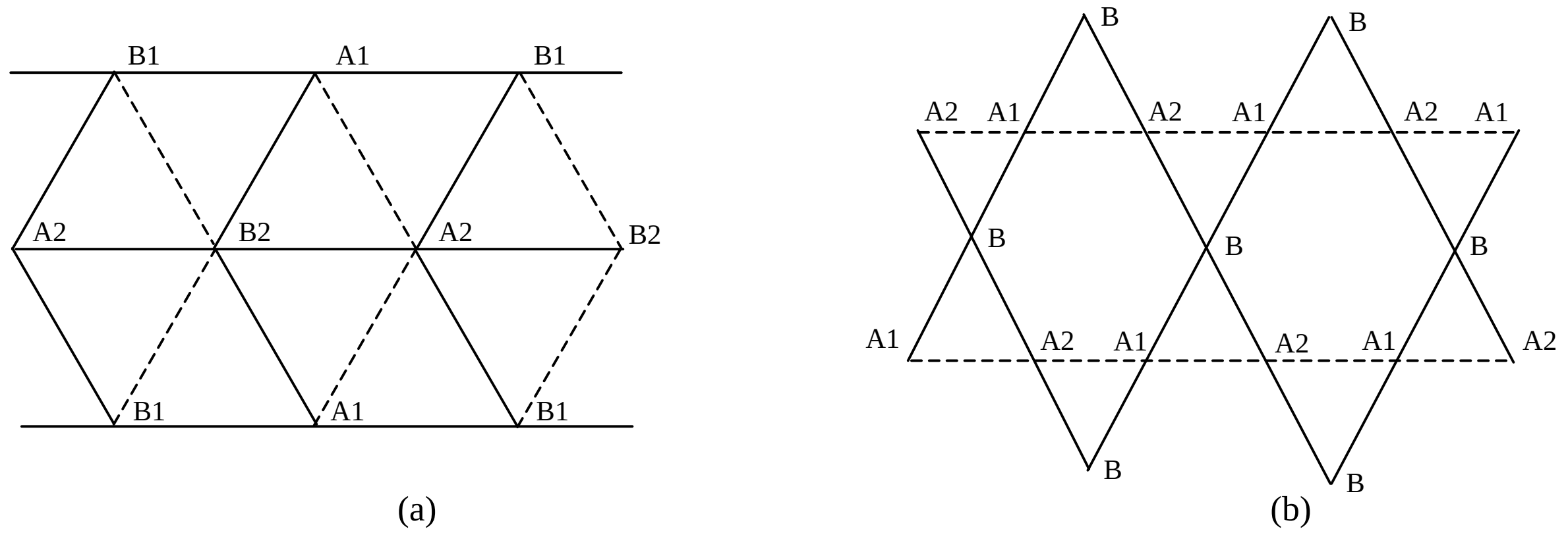

(a) (b)

Fig. 1: Possible collinear orderings in (a) triangular lattice and (b) kagomé lattice. Here A represents the up spin and B represents the down spin

Let us consider now the DM interaction along Y-direction. Then the classical energy is given by,

$$E_{Cl} = \frac{S^2}{\cos\phi}\left[\left(\cos(\phi-\theta_1)+\cos(\phi-\theta_2)+\cos(\phi-\theta_2+\theta_1)\right)\right] \qquad (3)$$

where, $\varphi = \tan^{-1}(D/J)$. The classical ground state is obtained after minimizing the above expression, and we have the condition

$$D\left[\alpha\cos 2\theta_2 - \alpha\sin 2\theta_2 - \cos\theta_2\right] = \sin\theta_2 \qquad (4)$$

with $\theta_1 = 2\theta_2$ for collinear (and spiral) order as mentioned above. It is easy to show that for $\alpha$ positive, $\theta_2$ cannot have values *0* and $\pi$ as solutions, for non-zero values of *D*. We can conclude, therefore, that the collinear order is not possible in two-dimensional topologically frustrated lattices in the presence of DM interaction for any value of α in the range $1 \geq \alpha > 0$.

Although there is no collinear order, Eqn. (4) admits of unique spiral solutions with two possible orientations: (i) $\theta_1 = 2\pi/3$, $\theta_2 = 4\pi/3$ for $D$ in the negative $Y$-direction and (ii) $\theta_1 = 4\pi/3$, $\theta_2 = 2\pi/3$ for $D$ in the positive $Y$-direction. This implies that the classical ground state is 120° Néel ordered and the direction of the DM interaction changes the chirality of the order. In Fig. 2 the two possible ground states are shown.

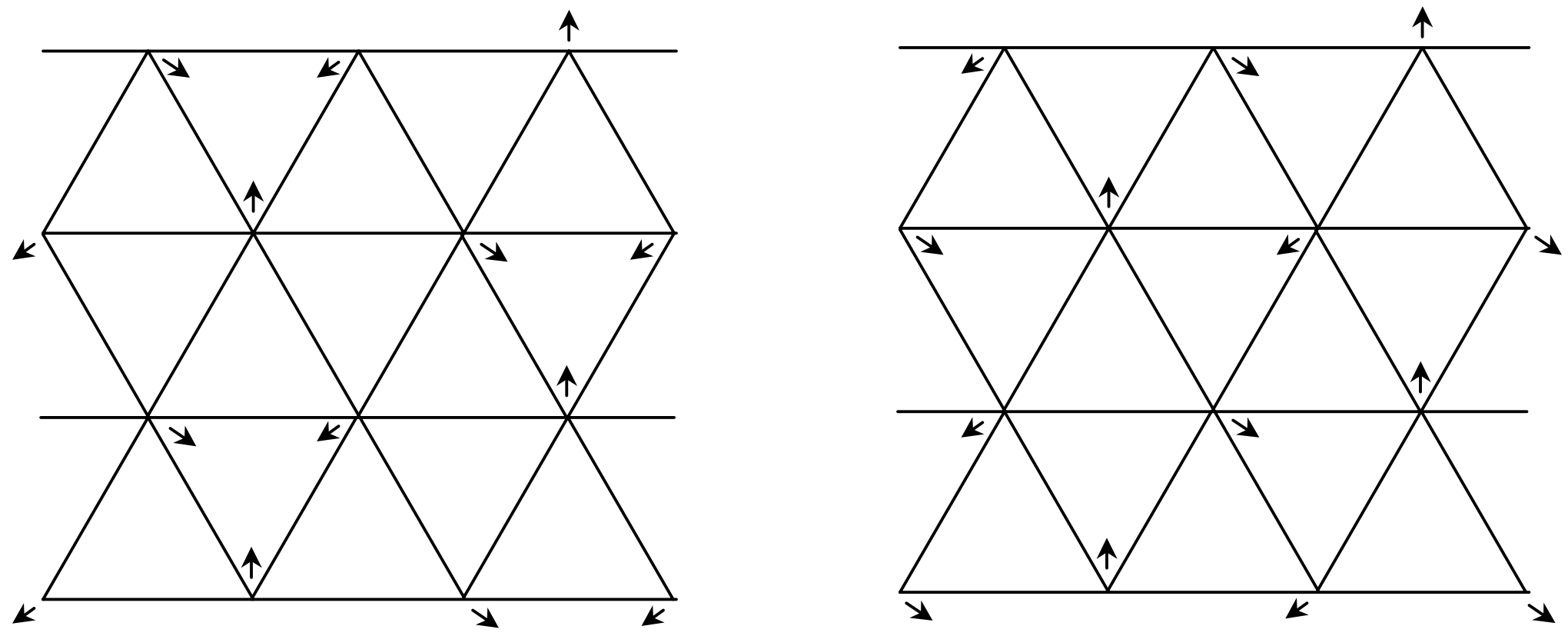

$D > 0$ D < 0

Fig. 2: Two possible Neél orderings for positive and negative values of DM interaction (acting perpendicular to the lattice plane) in a triangular lattice.

The case of the kagomé lattice is not so straightforward. A set of corner-sharing triangles generates the kagomé lattice. Harris, et. al. [23] had studied this lattice and described it by two types of configurations which they called $\mathbf{q} = 0$ (in which spins on each sublattices are parallel to each other and make an angle of 120° with the spins on the other two sublattices) and the $\sqrt{3}\times\sqrt{3}$ structure (in which one out of every four sites of an ordered triangular lattice is removed). In general, determination of the classical ground state in the kagomé lattice is very difficult, because of the large number of parameters. Recently, Elhajal, et. al.[23] studied the same Hamiltonian Eqn. (2) using Monté Carlo simulations. They observed that there is a possibility of two types of ground states in the $\mathbf{q}$=0 structure as shown in Fig. 3. Here too, the sign of DM interaction changes the chirality of the ground state as in the triangular lattice.

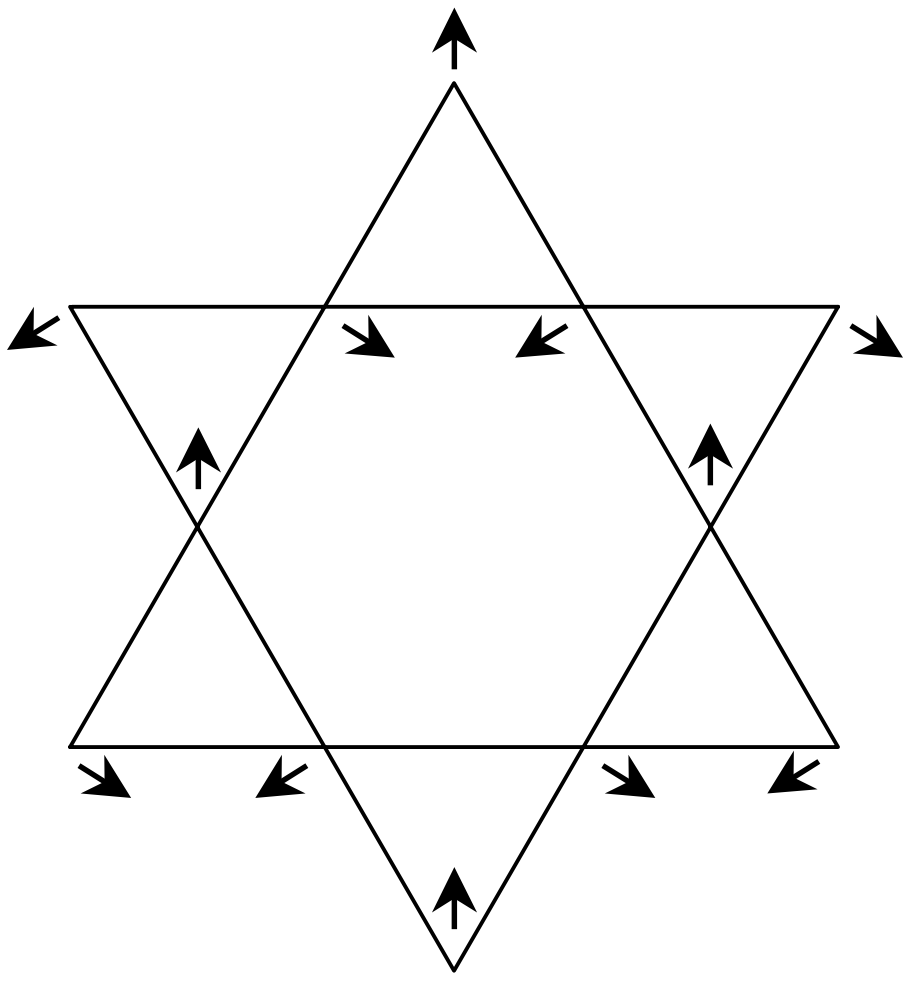

$D < 0$ $D > 0$

Fig. 3: Two possible Neél ordering in a kagomé lattice for positive and negative values of DM interaction acting perpendicular to the lattice plane.

## III. Linear Spin-Wave Analysis:

### (a) Collinear ordering

In this section we study the stability of collinear order in triangular and kagomé lattices in the HAFM $H = J\sum_{<i,j>}\vec{S}_i \cdot \vec{S}_j$. In the collinear ordering two possible spins are considered, A (up spins) and B (down spins). The usual Holstein-Primakoff (HP) transformation is, therefore

$$\begin{aligned} S^+_{A_j} &= \sqrt{2S}a_j \\ S^+_{B_j} &= \sqrt{2S}a^+_j \\ S^z_{A_j} &= S - a^+_j a_j \\ S^z_{B_j} &= -S + b^+_j b_j \end{aligned} \quad (5)$$

### (i) Triangular lattice:

A particular ordering is shown in Fig. 1. Here the dashed bonds are weaker bonds and allow a ferromagnetic ordering. We are considering four different kinds of spins, two for up spins A1 and A2 and two for down spins B1 and B2. After giving suitable HP transformations to introduce the four boson-operators, the Hamiltonian reads

$$H_k / J = -8NS^2 + 4\alpha NS^2 + S\left[\begin{pmatrix}\alpha^+_k & \beta_k\end{pmatrix} H_0(k) \begin{pmatrix}\alpha_k \\ \beta^+_k\end{pmatrix} - 2\right] \quad (6)$$

where $\alpha$ is the strength of the frustrated bond in the unit of other bonds.

$$H_0(k) = \begin{pmatrix} d & e_1^+ & e_3 & e_2 \\ e_1 & d & e_2 & e_3 \\ e_3^+ & e_2^+ & d & e_1 \\ e_2^+ & e_3^+ & e_1^+ & d \end{pmatrix} \quad (7)$$

and

$$\begin{aligned} d &= 4 - 2\alpha; \\ e_1 &= \alpha\left(e^{ik\cdot a_3} + e^{-ik\cdot a_2}\right); \\ e_2 &= e^{ik\cdot a_3} + e^{+ik\cdot a_2}; \\ e_3 &= 2e^{ik\cdot a_1}; \\ \alpha_k &= \begin{pmatrix} a_{1k} \\ a_{2k} \end{pmatrix} \\ \beta_k &= \begin{pmatrix} b_{1k} \\ b_{2k} \end{pmatrix} \end{aligned} \quad (8)$$

Here $r_1$ (1,0), $r_2$ (-1/2,√3/2) and $r_3$ (-1/2,-√3/2) are the bond directions through which the bosons are interacting. The excitation spectrum is shown in the Fig. 4(a). It is observed that this state is a possible ground state for values of α less than 0.15. Beyond this, the excitation spectrum becomes imaginary indicating that this order is destroyed by quantum fluctuations.

(ii) **Kagomé lattice**:

In this geometry, one can identify two types of A-spins (A1 and A2, say up spins) and one type of B-spin (down spin). These spin operators are transformed to bosonic operators through the Holstein-Primakoff transformations (Eq. 5) and the corresponding bosons are called $a_{1k}, a_{2k}, b_k$ respectively. These bosons are interacting through the bond directions $r_1(1,0), r_2(\frac{1}{2},-\frac{\sqrt{3}}{2}); r_3(\frac{1}{2},\frac{\sqrt{3}}{2})$. The expression for the Hamiltonian after Fourier transformation is

$$H \ / J = NS^2(-4+2\alpha) + S\sum_k \left[ \left(\alpha_k^+ \beta_k\right) H_0(k) \begin{pmatrix} \alpha_k \\ \beta_k^+ \end{pmatrix} - 4 + 2\alpha \right] \quad (9)$$

where,

$$H_0(k) = \begin{bmatrix} z_{1k} & z_{2k} \\ z_{2k} & z_{1k} \end{bmatrix}; \quad (10)$$

$$z_{1k}=\begin{pmatrix} 1-\alpha & \alpha\cos k\cdot\vec{r}_1 & 0 \\ \alpha\cos k\cdot\vec{r}_1 & 1-\alpha & 0 \\ 0 & 0 & 2 \end{pmatrix}$$

$$z_{2k}=\begin{pmatrix} 0 & 0 & e^{ik\cdot r_3} \\ 0 & 0 & e^{ik\cdot r_2} \\ e^{ik\cdot r_3} & e^{ik\cdot r_2} & 0 \end{pmatrix}$$

and

$$\alpha_k^+ = \begin{pmatrix} a_{1k}^+ \\ a_{2k}^+ \\ b_{-k}^+ \end{pmatrix}; \qquad (11)$$

$$\beta_k = \alpha_{-k}$$

The excitation spectrum is shown in Fig. 4(b). It is observed that the excitation spectrum remains real all over the Brillouin zone for $\alpha \le 0.11$ (less than the critical value for the triangular lattice) indicating the stability of LRO. The excitation spectrum in this case is flat in comparison to the triangular spectra, indicating that there may be other orderings close by in energy. Excitation of local soft modes discussed in the next section may destroy this particular collinear order easily.

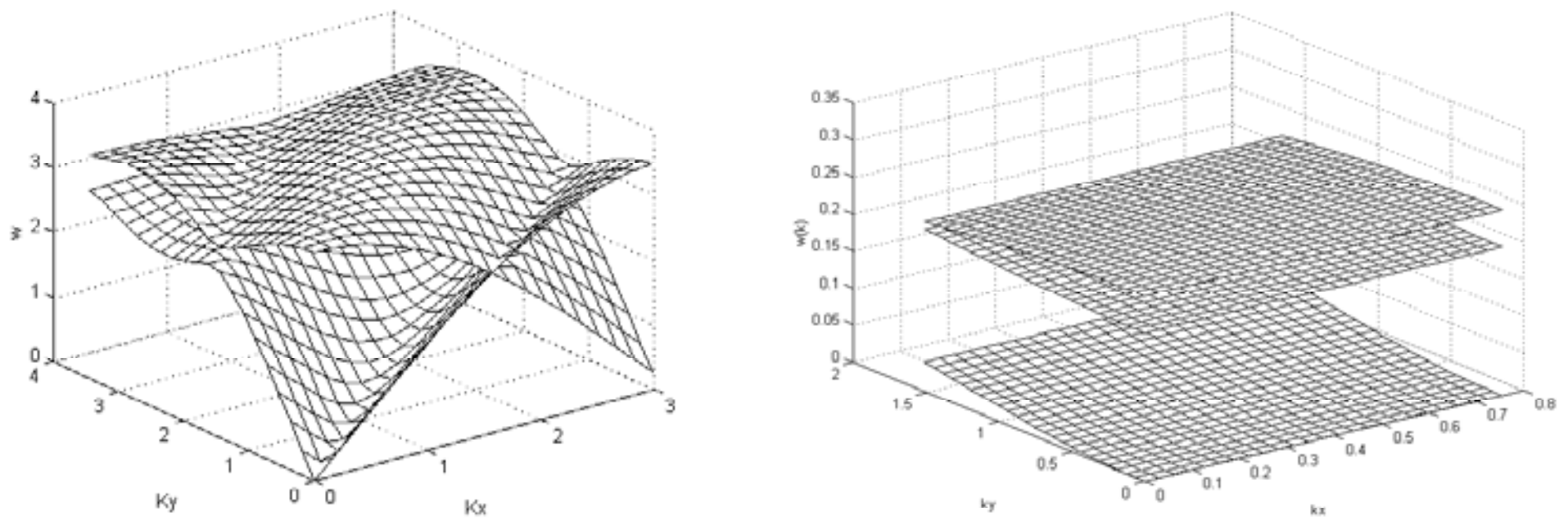

(a) (b)

Fig. 4. The excitation spectrum for collinear ordering. (a) for triangular lattice at $\alpha$ = 0.12 and (b) for Kagome lattice at $\alpha$ = 0.10

(b) **Spiral ordering**

In the presence of DM interaction, the collinear order disappears while the spiral order remains a candidate for the LRO state. We consider such canted spin-ordering presently, the spin vectors being oriented in the X-Z plane as before. To bring the neighbouring spins in the direction of the same magnetization axis (say Z-axis), one should rotate all the spins about Y-axis and in the new reference frame the spins are defined as,

$$S^{x}_{i}{}' = cosQ\, S^{x}_{i} + sinQ\, S^{z}_{i}\,;$$
$$S^{y}_{i}{}' = S^{y}_{i}\,; \qquad (12)$$
$$S^{z}_{i}{}' = -sinQ\, S^{x}_{i} + cosQ\, S^{z}_{i}\,;$$

Where, Q represents the angle of the spin vectors with respect to the Z-direction. In this new description we take the Holstein-Primakoff transformation, following the convention of ref. [9]

$$\mathrm{S}_{\mathrm{i}}^{\mathrm{x}} = \frac{1}{2}\sqrt{2S}(a_i + a_i^+)\,;\ \mathrm{S}_{\mathrm{i}}^{\mathrm{y}} = \frac{1}{2i}\sqrt{2S}(a_i - a_i^+)\ ;\mathrm{S}_{\mathrm{i}}^{\mathrm{z}} = (S - a_i^+ a_i) \qquad (13)$$

To study the excitation spectrum, we transform the bosons to Fourier space as usual $a_k = \frac{1}{\sqrt{N}}\sum_i e^{-i\vec{k}.\vec{r}_i} a_i$ where $N$ is the number of lattice points belonging to one sublattice. We consider the nearest-neighbor interaction only. Let $J_1$ and $D_1$ be the strength of nearest-neighbor Heisenberg interaction and nearest-neighbour DM interaction respectively. Thus we have finally the Hamiltonian

$$\begin{aligned} H = & \sum_{k,\alpha,\beta,i} S^2 (J_i \cos Q_{\alpha\beta} + D_i \sin Q_{\alpha\beta}) + hS \\ & - \frac{S}{2}(a^+_{\alpha k} a_{\alpha k} + a^+_{\beta k} a_{\beta k})(J_i \cos Q_{\alpha\beta} + D_i \sin Q_{\alpha\beta}) + h a^+_{\alpha k} a_{\alpha k} \\ & + \frac{S}{2}(a_{\alpha k} a_{-\beta k} e^{-\vec{i}k.r_i} + a^+_{\alpha k} a^+_{-\beta k} e^{\vec{i}k.r_i})(J_i \cos Q_{\alpha\beta} + D_i \sin Q_{\alpha\beta} - J_i) \\ & + \frac{S}{2}(a^+_{\alpha k} a_{\beta k} e^{\vec{i}k.r_i} + a_{\alpha k} a^+_{\beta k} e^{-\vec{i}k.r_i})(J_i \cos Q_{\alpha\beta} + D_i \sin Q_{\alpha\beta} + J_i) \end{aligned} \qquad (14)$$

($\alpha,\beta$) are the indices for the sublattices A, B and C. $Q_{\alpha\beta}$ is the angle between the spin vectors belonging to the sublattices $\alpha$ and $\beta$. The index $i$ above refers to the $i$-th neighbouring site. We derive $H_k$ in both triangular and kagomé lattice separately in the following.

(i) **Triangular lattice**

In the 120° Néel ordered state there are three types of spins, called A (Q=0), B (Q=120°) and C (Q=240°) belonging to three different sublattices. At first we consider the nearest neighbour interaction only. For $D_1$>0, $Q_{\alpha\beta}$ is taken as $4\pi/3$ and for $D_1$ <0, $Q_{\alpha\beta}$ is $2\pi/3$. In the following we consider the first case only. The ground states corresponding to both the chiralities have thus been taken into account. Every spin belonging to one particular sublattice interacts with the other through the bond directions $\boldsymbol{r_1}$ (1,0), $\boldsymbol{r_2}$ (-1/2,√3/2) and $\boldsymbol{r_3}$ (-1/2, -√3/2). We represent the bosons corresponding to the sublattices A, B and C by *a, b* and *c* and we have

$$H = \frac{9}{2}S^2(-J_1+\sqrt{3}D_1)+3hS+\frac{3S}{2}\sum_{\alpha,\beta,k}[(\alpha_k^+\beta_k)H_0(k)\begin{pmatrix}\alpha_k\\ \beta_k^+\end{pmatrix}-3C_1] \qquad (15)$$

where,

$$H_0(k)=\begin{bmatrix} M_1+M_2 & M_3\\ M_3 & M_1+M_2\end{bmatrix}; \qquad (16)$$

$$M_1=\begin{bmatrix} C_1 & 0 & 0\\ 0 & C_1 & 0\\ 0 & 0 & C_1\end{bmatrix};\ M_2=C_2\begin{bmatrix} 0 & z & z^*\\ z^* & 0 & z\\ z & z^* & 0\end{bmatrix};\ M_3=C_3\begin{bmatrix} 0 & z & z^*\\ z^* & 0 & z\\ z & z^* & 0\end{bmatrix}; \qquad (17)$$

with, $$z=\frac{1}{12}\left(e^{-ik.r_1}+e^{-ik.r_2}+e^{-ik.r_3}\right) \qquad (18)$$

$$\begin{aligned} C_1 &= J_1+\sqrt{3}D_1-\frac{h}{3S};\\ C_2 &= J_1+\sqrt{3}D_1;\\ C_3 &= -3J_1-\sqrt{3}D_1; \end{aligned} \qquad (19)$$

$$\alpha_k^+=\begin{pmatrix}a_k\\ b_k\\ c_k\end{pmatrix};\beta_k=\alpha_{-k}; \qquad (20)$$

The $6\times6$ matrix $H_0(k)$ can be diagonalized analytically by a general Bogoliubov transformation. It is interesting to note that all the 3×3 blocks building the matrix $H_0(k)$ are permutation matrices and thus can be diagonalised simultaneously in the basis

$$u_1=\begin{pmatrix}1 & 1 & 1\end{pmatrix};\ u_2=\begin{pmatrix}1 & j & j^2\end{pmatrix};\ u_3=\begin{pmatrix}1 & j^2 & j\end{pmatrix} \text{ and } j=e^{-2\pi i/3} \qquad (21)$$

The appearance of cubic roots of the unity is the manifestation of the ternary symmetry of the problem. We introduce the generalized Bogoliubov transformation as

$$\begin{bmatrix} A \\ B^+ \end{bmatrix} = T \begin{bmatrix} \alpha \\ \beta \end{bmatrix}. \qquad (22)$$

To preserve the boson commutation relations and to map $H_0(k)$ onto a diagonal matrix the column of the transformation matrix is written as $\begin{bmatrix} \lambda_i \, u_i \\ \mu_i \, u_i \end{bmatrix}$, where the coefficients $\lambda_i$ and $\mu_i$ satisfy the hyperbolic orthonormalization conditions $|\lambda_i|^2 - |\mu_i|^2 = 1$, etc. So once three column vectors are found, the remaining three vectors are obtained by the action of $-\sigma_x$, the first Pauli spin matrix. In this way the transformation matrix T is obtained. The diagonalised Hamiltonian then reads

$$H = \frac{9}{2}S^2(-J_1 - \sqrt{3}D_1) + 3hS + \frac{3S}{2}\sum_k \left[\psi_k^+ H_D \psi_k - 3C_1\right] \qquad (23)$$

or,

$$H = \frac{9}{2}S^2(-J_1 - \sqrt{3}D_1) + 3hS + 3S\sum_{k,L=1,2,3} \omega_L A_L^+ A_L + \frac{3S}{3}\sum_k (\omega_1 + \omega_2 + \omega_3 - 3C_1), \qquad (24)$$

where, $\omega_i = \sqrt{[C_1 + \rho_i(C_2 + C_3)][C_1 + \rho_i(C_2 - C_3)]}$ (25)

and $\rho_1 = z + z^*$; $\rho_2 = zj + z^* j^2$; $\rho_3 = z^* j + zj^2$. (26)

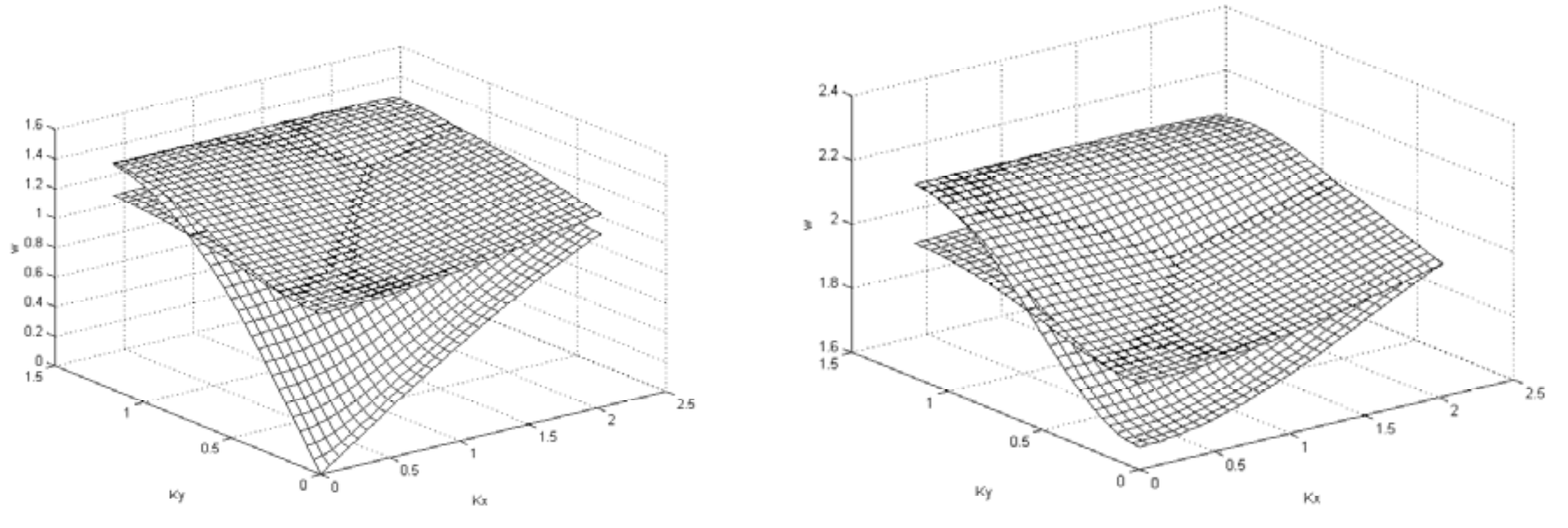

Fig. 5: Excitation spectra of the 120° order for the triangular lattice with $J_1 = 1$, $D_1 = 0.3$ for (a) h = 0 and (b) h= – 1

The excitation spectra are shown in Fig. 5. It is seen that the excitation spectrum is real and positive over the entire Brillouin zone. The excitation spectra have gaps at the centre of the Brillouin zone which are estimated as,

$$
\begin{aligned}
\Delta_{1(0,0)} &= \sqrt{(h/3S)\left(-3J_1+\sqrt{3}D_1-h/(3S)\right)} \\
\Delta_{2,3(0,0)} &= \sqrt{\left(h/3S+3\sqrt{3}/2D_1-3J_1/2\right)\left(\sqrt{3}D_1+h/(3S)\right)}
\end{aligned}
\tag{27}
$$

The ground state energy per bond including zero-point fluctuations is

$$
E_0=\frac{1}{2}S^2(-J_1-\sqrt{3}D_1)+\frac{hS}{3}+\frac{S}{2N}\sum_k(\omega_1+\omega_2+\omega_3)+\frac{S}{2}(-J_1-\sqrt{3}D_1+\frac{h}{3}) \tag{28}
$$

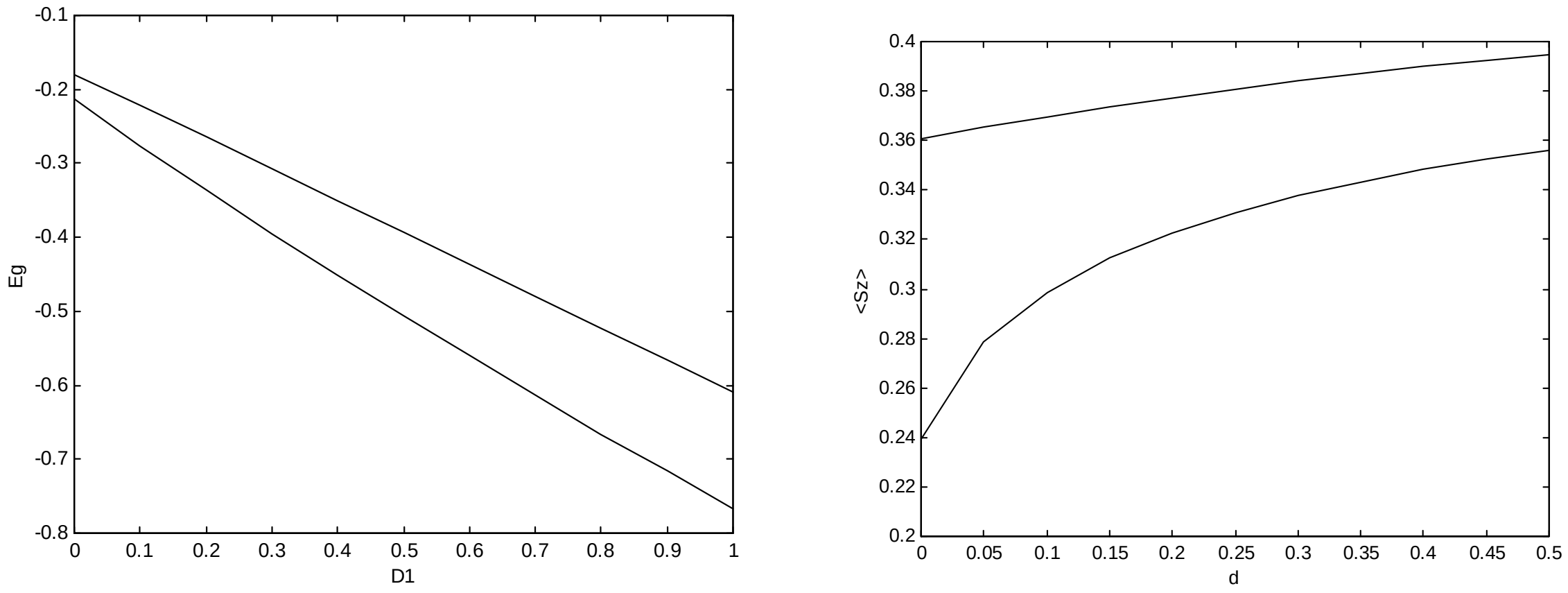

(a) (b)

Fig. 6: (a) Variation of the ground state energy for the triangular lattice w.r.t $D_1$ at h = 0 (upper) and h = – 1.0 (lower); (b) The sublattice magnetization versus $D_1$ at h=0 (lower) and h=-0.3 (upper) for $J_1$=1 and S=1/2.

In the above expression the last two terms give the quantum correction to the classical ground state energy. In Fig. 6(a) we plot the ground state energy as a function of $D_1$ and it is observed that the total ground state energy decreases as $D_1$ increases. There is also a reduction in the ground state sublattice magnetization due to quantum fluctuations which is calculated per site as

$$
m=-\frac{1}{2}+\frac{1}{2N}\sum_{i,k}\frac{C_1+C_2\rho_i}{\omega_i(k)} \;. \tag{29}
$$

In Fig. 6(b) we plot the sublattice magnetization ($\langle S_z\rangle=S-m;\ S=1/2$) as a function of $D_1$ and this also increases as $D_1$ increases. If we introduce the second-neighbour interaction (spin-spin interaction within the same sublattice) we have the Hamiltonian in the same form with modified constants as

$$C_1^{/} = +J_1 + \sqrt{3}D_1 - \frac{h}{3S} - 2J_2 + \frac{2J_2}{3}\sum \cos \vec{k}.\vec{s}_i;$$

$$C_2^{/} = J_1 - \sqrt{3}D_1; \qquad (30)$$

$$C_3^{/} = -3J_1 - \sqrt{3}D_1;$$

where, $\boldsymbol{s_1}$=$\boldsymbol{r_3} - \boldsymbol{r_2}$; $\boldsymbol{s_2}$=$\boldsymbol{r_1} + \boldsymbol{r_3}$ ; $\boldsymbol{s_3} = \boldsymbol{r_1} + \boldsymbol{r_2}$.

The corresponding excitation spectra are shown in the Fig. 7. The nature of the excitation spectrum remains similar to Fig. 5.

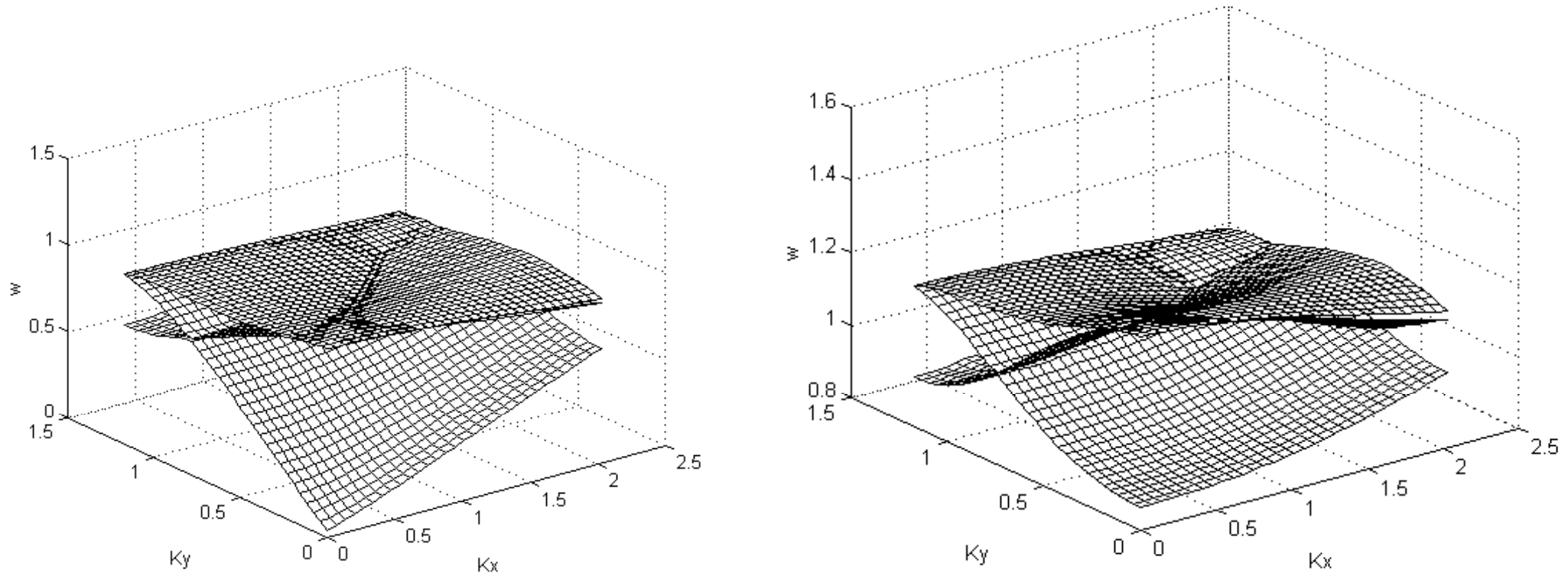

(a) (b)

Fig. 7. The excitation spectra for 120°ordering on a triangular lattice with $D_1$=0.5, $J_1$=1and $J_2$=0.2 (a) for h=0 and (b) h=–0.3

(ii) **Kagomé lattice**

It has been observed from the results of Monté Carlo simulation that there is a possibility of two types of ground state (**q**=0) in the kagomé lattice depending upon the direction of the DM interaction. Three types of sublattices, A, B and C describe the spins of the lattice. Each spin interacts with the spins of the other sublattice via the nearest neighbor Heisenberg (of strength $J_1$) and DM (of strength $D_1$) interactions along the bond directions $\vec{r}_1(1,0)$, $\vec{r}_2(\frac{1}{2},-\frac{\sqrt{3}}{2})$, $\vec{r}_3(\frac{1}{2},\frac{\sqrt{3}}{2})$. For one of the ground states (say, $D_1$ positive) $Q_{\alpha\beta}$ is $2\pi/3$ or $4\pi/3$ depending on the chirality of the nearest neighbor spin configuration and for the other ground state ($D_1$ negative) they are interchanged. So ultimately we have

$$H = C + \frac{S}{4}\sum_{\alpha,\beta,k}[(\alpha_k^{+}\beta_k)H_0(k)\begin{pmatrix}\alpha_k \\ \beta_k^{+}\end{pmatrix} - d] \qquad (31)$$

where, $$\alpha_k = \begin{pmatrix} a_k \\ b_k \\ c_k \end{pmatrix}; \beta_k = \alpha_{-k}. \qquad (32)$$

The new Hamiltonian to be diagonalized is

$$H_0(k) = \begin{bmatrix} M_1 & M_2 \\ M_2 & M_1 \end{bmatrix}; \quad (33)$$

where, $$M_1 = \begin{bmatrix} d & e_3^+ & e_2 \\ e_3 & d & e_1^+ \\ e_2 & e_1 & d \end{bmatrix}; \quad (34)$$

$$M_2 = \begin{bmatrix} 0 & f_3^+ & f_2 \\ f_3 & 0 & f_1^+ \\ f_2 & f_1 & 0 \end{bmatrix}. \quad (35)$$

and $$C = 3S\sum_k J_1 S - 2h \quad (36)$$

$$d = -12J_1 + 12h + \frac{12h}{S} \quad (37)$$

$$e_p = J_1 \cos(k\cdot r_p) + \sqrt{3}D_1 i \sin(k\cdot r_p);$$
$$f_p = -3J_1 \cos(k\cdot r_p) + \sqrt{3}D_1 i \sin(k\cdot r_p).$$

This Hamiltonian cannot be diagonalized analytically. Preserving the bosonic commutation relation the above Hamiltonian is diagonalized numerically. It is observed that one of the eigenfrequencies in the spin-wave spectrum remains nondispersive**.** This dispersionless mode is due to the excitation of zero-energy local modes. Classically, for spins lying in the *X-Z* plane, these modes correspond to tipping of spins alternately in and out of the plane. If the tipping angle is $\theta$, the energy of such a mode is proportional to $\theta^4$ with a positive coefficient. Thus in the classical case, anharmonicity apparently stabilizes the soft modes, and presumably this will also be the case for the quantum spins.

Adding the second-neighbor interaction (of strength $J_2$ and $D_2$; the bond directions being $\boldsymbol{t_1}=\boldsymbol{r_3} - \boldsymbol{r_2}$; $\boldsymbol{t_2}=\boldsymbol{r_1} + \boldsymbol{r_3}$ ; $\boldsymbol{t_3} = \boldsymbol{r_1} + \boldsymbol{r_2}$ ) and third-neighbor interaction (of strength $J_3$ ; the bond directions being $\boldsymbol{s_1}=2\boldsymbol{r_1}$; $\boldsymbol{s_2}=2\boldsymbol{r_2}$ ; $\boldsymbol{t_3} = 2\boldsymbol{r_1} - 2\boldsymbol{r_2}$ ) lead to some interesting results. It is interesting to note that there are two inequivalent sets of third neighbours, one is obtained by two nearest neighbor steps and the other (on opposite sides of the hexagon), by three such steps. Following Harris, et al. we also include $J_3$ only (the first type) and the latter coupling is neglected. In the third- neighbor term there is no effect from DM interaction because the interaction is among spins in the same sublattice. Then above expressions (34-37) are redefined as

$$M_1 = \begin{bmatrix} d_1 & e_3^+ & e_2 \\ e_3 & d_2 & e_1^+ \\ e_2 & e_1 & d_3 \end{bmatrix}; \quad (38)$$

$$M_2 = \begin{bmatrix} 0 & f_3^+ & f_2 \\ f_3 & 0 & f_1^+ \\ f_2 & f_1 & 0 \end{bmatrix} \quad (39)$$

$$C = -3\sum_{k} S^2 (J_1 + J_2 - 3J_3) \tag{40}$$

$$d = -12\left(J_1 + J_2 - 2J_3\right) + 12h + \frac{12h}{S}$$

$$d_1 = 4(J_1 + J_2) - 8J_3 (1 - \frac{1}{2}[\cos(k \cdot s_2) + \cos(k \cdot s_3)]$$

$$d_2 = 4(J_1 + J_2) - 8J_3 (1 - \frac{1}{2}[\cos(k \cdot s_3) + \cos(k \cdot s_1)] \tag{41}$$

$$d_3 = 4(J_1 + J_2) - 8J_3 (1 - \frac{1}{2}[\cos(k \cdot s_1) + \cos(k \cdot s_2)]$$

$$e_p = J_1 \cos(k \cdot r_p) + J_2 \cos(k \cdot t_p) + \sqrt{3} D_1 i \sin(k \cdot r_p) + \sqrt{3} D_2 i \sin(k \cdot t_p);$$

$$f_p = -3(J_1 \cos(k \cdot r_p) + J_2 \cos(k \cdot t_p)) + \sqrt{3} i (D_1 (\sin(k \cdot r_p) + D_2 (\sin(k \cdot t_p)). \tag{42}$$

$$p = 1,2,3$$

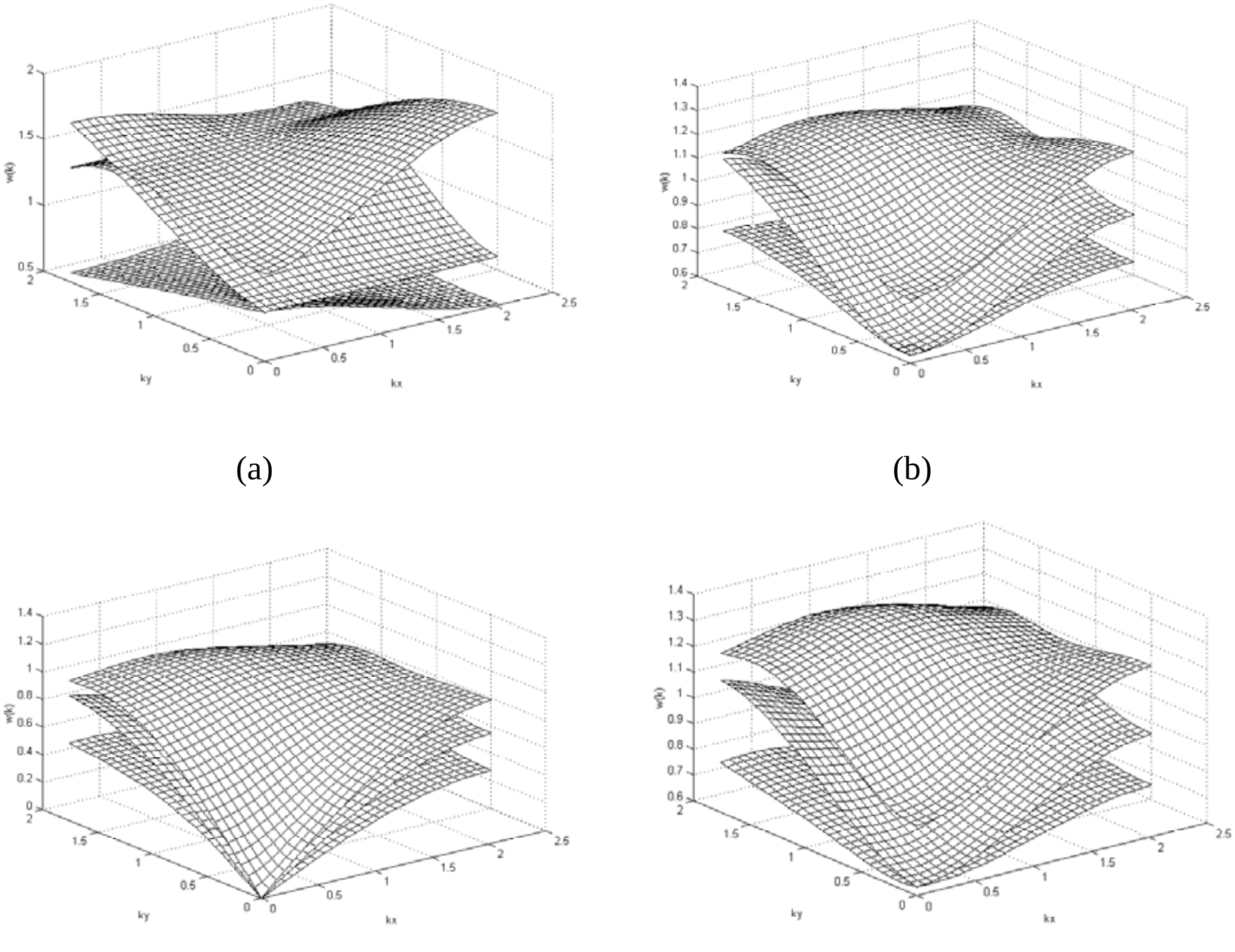

Fig. 8(a) Excitation spectra for the kagomé lattice with 120°ordering, $J_1$= 1, $J_2$ = 0.2, $J_3$ = 0.1, $D_1$= 0.8, h = – 0.2; (b) Same with $J_1$= 1, $J_2$ = 0.2, $J_3$ = 0.1,$D_1$= $D_2$= 0 & h = – 0.1, (c) for $J_1$ = 1, $J_2$ = 0.2, $J_3$ = 0.1, $D_1$ = $D_2$ = 0.2 & h = 0; and (d) for $J_1$ = 1, $J_2$ = 0.2, $J_3$ = 0.1, $D_1$ = $D_2$ = 0.2, h = – 0.1.

The excitation spectra have been obtained by numerical diagonalization of the above matrix preserving their commutation relations. In Fig. 8 we plot the excitation spectrum for the kagomé lattice. It is observed that all the three branches of each spectrum are dispersive and have gaps.

The ground state energy per site can be calculated from the expression,

$$\frac{E_0}{3N} = -S(S+1)\left(J_1 + J_2 - 2J_3\right) - h(S-1) + \frac{S}{8}\left(\frac{1}{3N}\sum \omega_1 + \omega_2 + \omega_3\right) \quad (43)$$

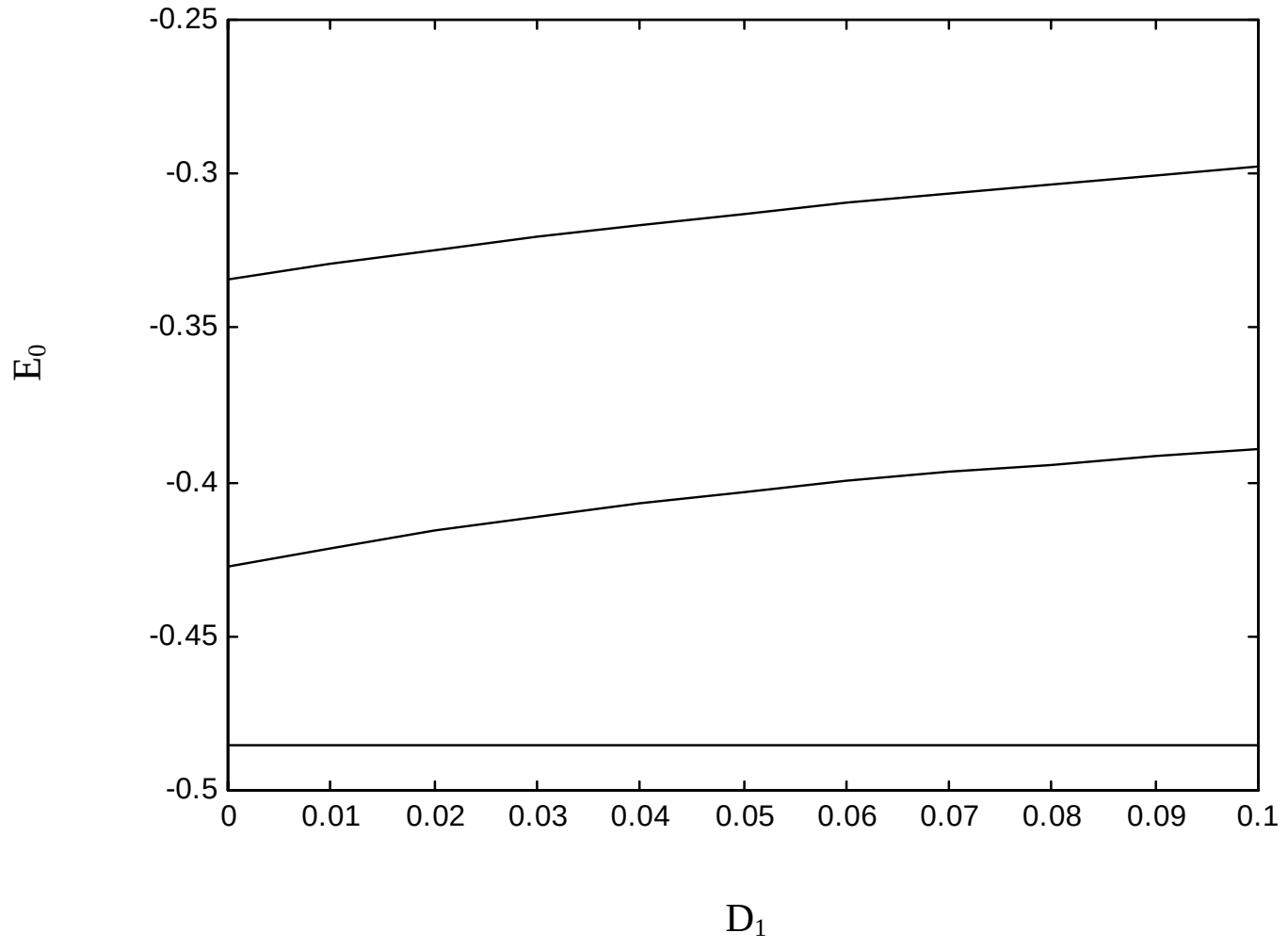

(a)

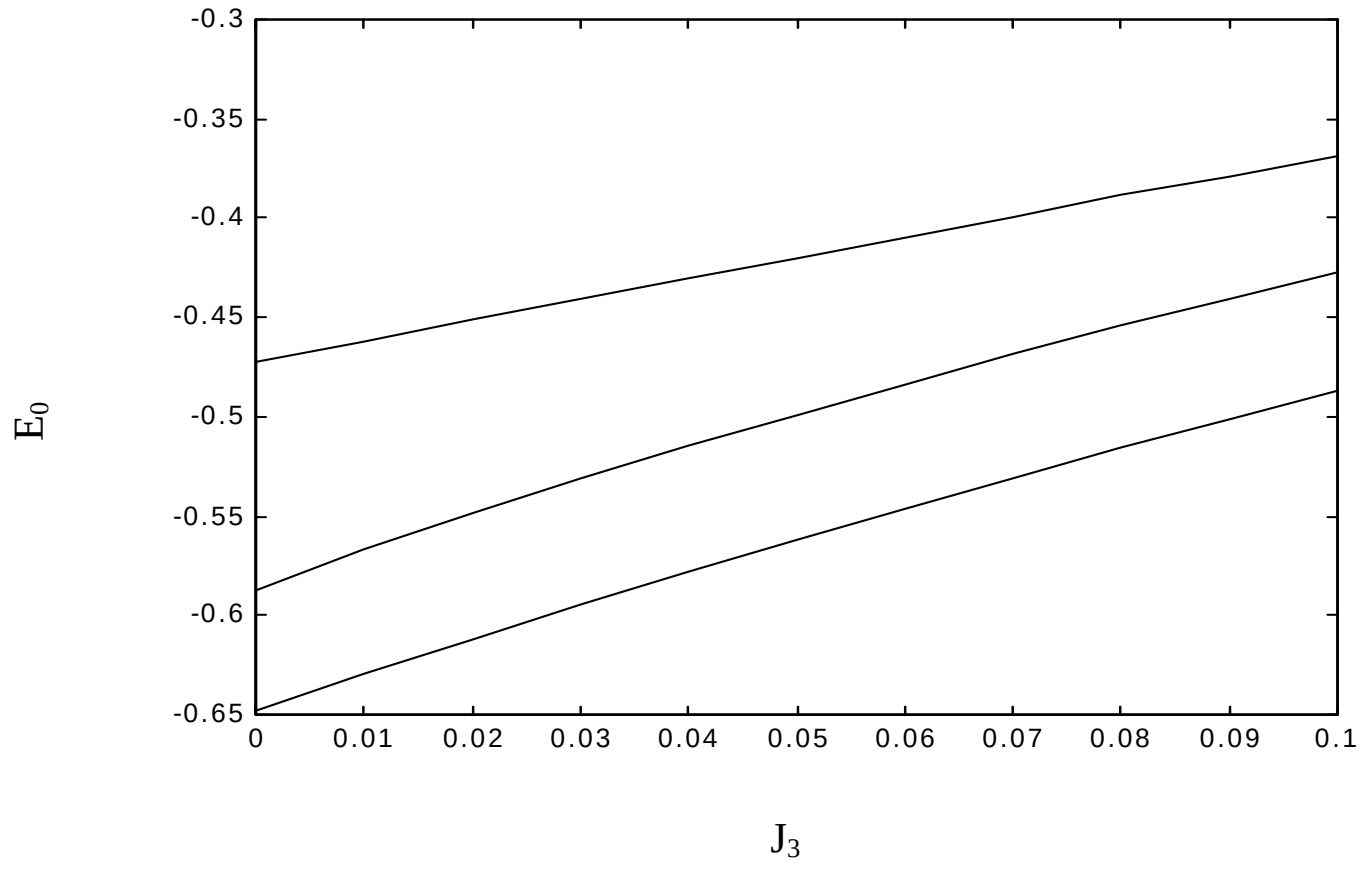

(b)

Fig. 9. Variation of the ground state energy for kagomé lattice (a) with $D_1$ for $J_1$=1; $J_2$=0.1; $J_3$=0.05; $D_2$=0.1; h=0, -0.1, -.2 (upper to lower). (b) with $J_3$ for $J_1$=1; $D_1$=0; $D_2$=0.0; h=0, $J_2$=0, 0.1, 0.2 (upper to lower).

In Fig. 9 we show how the ground state energy varies as the Heisenberg and the DM interaction strengths are varied.

## V. Conclusion

To summarize, in this paper we have considered the possibility of long-range order in two geometrically frustrated lattices in two dimensions, namely the kagomé and triangular lattices, in the isotropic and anisotropic Heisenberg Hamiltonian. It is observed that collinear ordering is possible in these lattices if one of the three bonds in each triangular plaquette is allowed to be frustrated. From linear spin-wane analysis a small parameter regime is identified where the collinearly ordered state is the possible ground state. Quantum fluctuation destroys such collinear order beyond $\alpha$ = 0.15 (0.11) for the triangular (kagomé) lattice. It has also been observed that anisotropic DM interaction disfavours collinear order in the ground state. We show that the Heisenberg Hamiltonian in the presence of out of plane DM interaction and in-plane magnetic field leads to a possible ground state with spiral ordering (angle 120º between the spins). There are states with different chiralities depending upon the direction of the DM interaction. The excitation spectrum has been obtained analytically for nearest-neighbour Heisenberg Hamiltonian in the triangular lattice and it is found to have two degenerate modes. The magnetic field or further-neighbour interaction removes the degeneracy. The effect of DM interaction is to introduce a gap in one of the branches of the excitation spectrum while the magnetic field, as expected, introduces a gap in all the branches. We also calculated the ground state energy and the sublattice magnetization analytically for the nearest-neighbour model with DM interaction and the magnetic field, and we observe that the ground state energy decreases and the sublattice magnetization increases as the DM interaction increases. This implies that LRO stabilizes with the increase of the strength of the DM interaction in these lattices.

The nearest-neighbour Heisenberg Hamiltonian on the kagomé lattice has a dispersionless branch in the excitation spectrum indicating the presence of soft modes, which destabilize the 120º ordering in its ground state. Inclusion of further-neighbour interactions stabilizes LRO as the flat band becomes dispersive. We have seen that the DM interaction in the presence of magnetic field stabilizes the soft modes in the near-neighbour interaction regime. Further-neighbour Heisenberg model along with the DM interaction have also been taken into account. In that case, the dispersionless branch disappears and a gap opens up in the excitation spectrum as the magnetic field is applied. An estimate of the ground state energy is made for different values of further-neighbour interactions and the DM interactions. Our calculations therefore imply that although Heisenberg antiferromagnetic model involves disordered states in two dimensional frustrated lattices, there are possible long-range ordered ground states in models that include anisotropic and further-neighbour interactions.

## Ackowledgement

The work is supported by UGC minor research scheme (Letter No: F. PSW-057/04-05(ERO)). We are thankful to Y. Lee and M. Elhajal for sending us ref. 24 and unpublished result (ref. 20) respectively.

## References

*umabhaumik@lycos.com
**arghya@phy.iitkgp.ernet.in, arghya@mpipks-dresden.mpg.de